\documentclass[11pt]{article}

\usepackage[T1]{fontenc}
\usepackage[utf8]{inputenc}
\usepackage{authblk}

\usepackage{upgreek}
\usepackage{hhline}
\usepackage{multirow}
\usepackage{amsmath,amsfonts,amssymb}
\usepackage{booktabs}
\usepackage{makecell}
\usepackage{graphicx}

\usepackage[
    backend=biber,
    style=numeric-comp,
    sorting=none,
    doi=true,
    isbn=false,
    url=false,
    eprint=false,
    date=year,
    maxbibnames=99,
    maxcitenames=99
]{biblatex}
\renewbibmacro{in:}{}

\DeclareNameFormat{default}{%
  \usebibmacro{name:family-given}
    {\namepartfamily}{\namepartgiveni}{\namepartprefix}{\namepartsuffix}%
  \usebibmacro{name:andothers}}

\DeclareFieldFormat[article]{title}{#1}

\newbibmacro*{journalname}{%
  \iffieldundef{shortjournal}
    {\printfield{journaltitle}}
    {\printfield[journaltitle]{shortjournal}}}
\DeclareFieldFormat{journaltitle}{\mkbibemph{#1}}

\DeclareFieldFormat[article,periodical,suppperiodical]{volume}{\mkbibbold{#1}}
\AtEveryBibitem{\clearfield{number}}

\DeclareFieldFormat{pages}{#1}

\DeclareFieldFormat{doi}{(doi:\space\seqsplit{#1})}

\DeclareBibliographyDriver{article}{%
  \usebibmacro{bibindex}%
  \usebibmacro{begentry}%
  \usebibmacro{author}%
  \setunit{\addspace}%
  \printfield{year}%
  \setunit{\addspace}%
  \iffieldundef{title}{}{\printfield{title}\setunit{\addspace}}%
  \usebibmacro{journalname}%
  \setunit{\addspace}%
  \printfield{volume}%
  \setunit{\addspace}%
  \printfield{pages}%
  \setunit{\addspace}%
  \printfield{doi}%
  \usebibmacro{finentry}}

\usepackage[hidelinks]{hyperref}
\usepackage[margin=2.5cm]{geometry}
\usepackage{seqsplit}

\title{Design and simulation of highly selective graphene-silicon nitride
integrated dual-mode electro-absorption modulators}

\author[1,*]{Fernando Martín-Romero}
\author[1]{Víctor J. Gómez}

\affil[1]{Nanophotonics Technology Center, Universitat Politècnica de
València, Valencia, 46022, Spain}
\affil[*]{fmarrom@ntc.upv.es}

\date{}

\begin{document}

\maketitle

\begin{abstract}
We present the design, simulation and optimization of an integrated graphene-silicon nitride dual-mode modulator with high mode selectivity, optimized for a wavelength of 1.55 $\mathrm{\upmu}$m. The proposed device finds applications in mode division multiplexing systems, enabling the simultaneous and independent switching of two transverse electric modes supported by a buried dual-mode silicon nitride waveguide (TE$_0$ and TE$_1$). Three graphene-Al$_2$O$_3$-graphene nanoribbons are integrated on top of the waveguide. A central nanoribbon acts as a TE$_0$ mode absorber, while the two remaining nanoribbons located on the sides act as TE$_1$ mode absorbers. Their absorption is tuned by modifying the Fermi energy of graphene, which can be achieved through electrical doping. Thus, the corresponding nanoribbons operate as selective TE$_0$ and TE$_1$ modulators with a total energy consumption per unit length under 2430 pJ bit$^{-1}$ cm$^{-1}$. The modulation depth of each mode can reach up to 316 dB/cm and 273 dB/cm, respectively, while maintaining a selection ratio of 5.63--6.28 and extinction ratio of 204--248 dB/cm between them. Under thinner Al$_2$O$_3$ conditions, consumption below 607.5 pJ bit$^{-1}$ cm$^{-1}$ can be achieved with a marginal reduction of the optical performance in the TE$_1$ modulator. Overall, we report the design of a fully optimized integrated dual-mode modulator based on graphene tunable electro-absorption, opening the path toward integrated efficient multimode optical communication systems.
\end{abstract}

\section{Introduction}

Graphene-silicon nitride integrated waveguides provide a highly promising platform for the realization of efficient graphene-based dual-mode modulators \cite{martin-romero_design_2024b}. Within all the unique properties of graphene \cite{bonaccorso_graphene_2010}, the tunability of its broadband optical response through electrical gating and its ultra-high carrier mobility have led to a wide range of applications, such as optical isolators \cite{ma_hybrid_2016, zarei_tunable_2021}, modulators \cite{ono_ultrafast_2020, liu_doublelayer_2012a, hu_broadband_2016, koester_highspeed_2012, heidari_integrated_2022}, mode filters \cite{martin-romero_design_2026b}, sensors \cite{redolat_hybrid_2026}, and photodetectors \cite{ma_plasmonically_2019, gao_grapheneonsilicon_2018}. The normal incidence of light in a graphene monolayer results in a broadband optical absorption of only 2.3\% \cite{nair_fine_2008}. However, that proportion can be increased by employing hybrid integration. When graphene is transferred on top of a waveguide, the in-plane evanescent field of the guided modes couples to it. As a consequence, total absorption or phase change in the transmitted signal is enhanced \cite{koester_waveguidecoupled_2014a}. Thus, the combination of the distinctive properties of graphene with the well-established silicon and silicon nitride (Si$_3$N$_4$) platforms has resulted in multiple proposals and demonstrations of novel integrated graphene hybrid devices such as high-speed broadband photodetectors \cite{gao_highperformance_2018}, all-optical switches \cite{ono_ultrafast_2020, qiu_highperformance_2021}, robust athermal modulators \cite{dalir_athermal_2016} or ultra-fast electro-optical modulators \cite{liu_doublelayer_2012a, hu_broadband_2016, koester_highspeed_2012, heidari_integrated_2022}.

\begin{figure}[ht!]
\centering
\includegraphics[width=\linewidth]{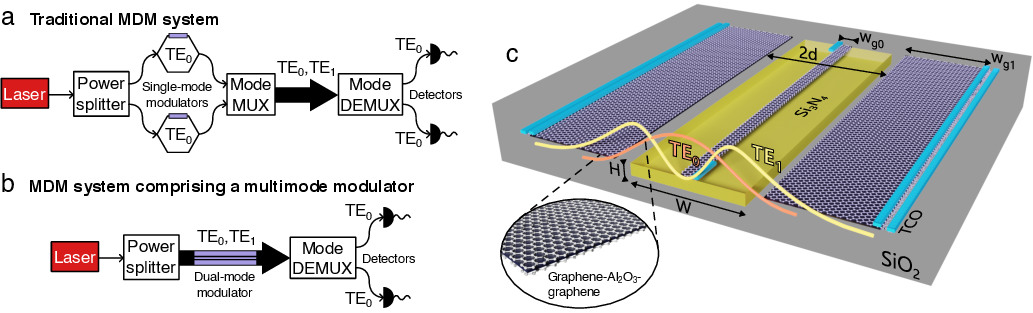}
\caption{(a) MDM communication system comprising single-mode modulation. (b) MDM communication system comprising dual-mode modulation. (c) Schematic view of the proposed graphene-Si$_3$N$_4$ integrated dual-mode electro-absorption modulator including graphene-Al$_2$O$_3$-graphene capacitor nanoribbons.}
\label{fig:1}
\end{figure}

The present study focuses on integrated multimode modulators, which allow the simultaneous and independent modulation of multiple spatial modes within a waveguide \cite{martin-romero_design_2024b, yue_graphenebased_2020a, wang_design_2021, lian_modeselective_2022a}. They can be implemented in multimode integrated optical communications. Their primary application lies in mode division multiplexing (MDM) systems \cite{stern_onchip_2015, jia_optical_2017, liu_arbitrarily_2019}, which utilize higher order modes supported by a multimode waveguide to enhance the capacity of a single-wavelength channel. In a traditional MDM system, the power emitted by the laser source is distributed among an array of single-mode modulators \cite{dai_silicon_2013} (figure \ref{fig:1}(a)). Then, the modulated signals are converted into higher order guided modes and multiplexed into a multimode waveguide. Such a configuration induces very high insertion losses and a large footprint. Replacing the array of single-mode modulators with one multimode modulator would allow a more compact and efficient design, helping to overcome the cited disadvantages (figure \ref{fig:1}(b)). However, the inclusion of highly efficient multimode modulators in MDM systems still remains a challenge.

In this work, we present the design and simulation of an integrated graphene-Si$_3$N$_4$ dual-mode electro-absorption modulator optimized for a wavelength of 1550 nm. The optimization of the device is preceded by a computational study in which TE$_0$ and TE$_1$ mode-selective modulation is investigated through simulations based on active tuning of integrated graphene passive mode filters \cite{martin-romero_design_2024b, martin-romero_design_2026b, xing_waveguideintegrated_2019a, xing_ultrabroadband_2018a}. The mode filters consist of a dual-mode silicon or Si$_3$N$_4$ waveguide, on top of which graphene nanoribbons are transferred. Tailoring the position and dimensions of the nanoribbons allows them to act as mode-selective absorbers. A nanoribbon located over the center of the waveguide core presents a significantly stronger absorption of the TE$_0$ mode with respect to the TE$_1$ mode. Similarly, two nanoribbons located symmetrically on the sides of the waveguide induce a stronger absorption for the TE$_1$ mode than for the TE$_0$ mode. Owing to the tunable absorption of graphene \cite{wang_gatevariable_2008}, which can be achieved by electrical gating, mode filters can be converted into tunable filters or mode-selective modulators. Thus, low-loss transparent conductive oxide (TCO) electrodes can be connected to the nanoribbons in order to develop TE$_0$ or TE$_1$ modulators. We report on the influence of the core waveguide material, Fermi energy, and charge relaxation time of graphene on the extinction, selection ratio, and modulation depth of the mode-selective modulators. Combining the TE$_0$ and TE$_1$ modulators on one waveguide and optimizing the nanoribbon geometry, we design a device capable of switching between the four logical states corresponding to the transmission of one, both or none of the modes: (1,0), (0,1), (1,1) and (0,0). A schematic of the full device is shown in figure \ref{fig:1}(c), comprising dual-layer graphene nanoribbons with Al$_2$O$_3$ as a dielectric spacer. The TE$_0$ mode selectivity is enabled by the overlap of the central nanoribbons with the maximum intensity of the TE$_0$ mode and the minimum intensity of the TE$_1$ mode. Similarly, the deeper evanescent tail of the TE$_1$ mode is overlapping more strongly with the side nanoribbons, enabling TE$_1$ mode selectivity. We compute the range of values achieved for the modulation depth of both modes, as well as the required gate voltage and energy consumption of the device. The subsequent sections outline the simulation framework and report the corresponding results.

\section{Mode-selective modulators}

\subsection{Simulation framework}

The efficiency of the mode-selective modulators was assessed by analyzing the propagation losses induced by the nanoribbons in the TE$_0$ and TE$_1$ modes at a wavelength of $1550$ nm. Graphene was modeled as a two-dimensional material with surface conductivity $\sigma_g$, including two contributions corresponding to intraband and interband absorption processes. Both were derived from the Kubo formula \cite{andryieuski_graphene_2013} as

\begin{gather}
    \sigma_{intra}=\frac{2 k_B T e^2}{\pi \hbar^2} \mathrm{ln}\left(2  \mathrm{cosh} \frac{E_f}{2 k_B T}\right) \frac{i}{\omega+i \tau^{-1}}, \\
    \sigma_{inter}=\frac{e^2}{4 \hbar} \left( H(\omega/2)+i\frac{4\omega}{\pi} \int_0^{\infty} \frac{H(\Omega)-H(\omega/2)}{\omega^2-4\Omega^2} \mathrm{d}\Omega \right), \\
    \sigma_g=\sigma_{intra}+\sigma_{inter},
\end{gather}

where $\tau$ is the charge relaxation time, $T$ is the temperature, $\omega=2 \pi f$ is the angular optical frequency, $E_f$ is the Fermi energy and $H(\Omega)=\mathrm{sinh}(\frac{\hbar \Omega}{k_B T})/[\mathrm{cosh}(\frac{\hbar \Omega}{k_B T})+\mathrm{cosh}(\frac{E_F}{k_B T})]$. A temperature of 300 K was considered in the simulations.

When the Fermi energy is $E_f>(\hbar\omega)/2$, the conductivity of graphene experiences a transition from opaque to transparent behavior. In the opaque regime, the interband processes are responsible of a stronger optical absorption. Beyond the discussed threshold, the intraband processes are the main source of absorption. However, their low strength causes the optical absorption to decrease sharply. For the wavelength of 1550 nm, this threshold is $E_f\approx0.4$ eV \cite{koester_waveguidecoupled_2014a}. Electrical tuning of $E_f$ is a mechanism that allows the modulation of optical absorption. In the absence of electrical doping, we assumed a negligible natural doping of $E_f\approx0$. A standard charge relaxation time of $\tau=10^{-13}$ s was selected for the majority of the simulations \cite{andryieuski_graphene_2013}. As indicated by the Kubo formula, this last parameter only impacts the conductivity of intraband processes. Thus, its influence on the optical absorption of undoped graphene is residual.

The TE$_0$ and TE$_1$ modes were computed using the finite element method (FEM) through COMSOL Multiphysics\textregistered \ mode analysis. The single-layer graphene nanoribbons have a thickness of  $\approx1$ nm \cite{martin-romero_design_2026b}. Thus, their effect on the mode profiles is not significant and was excluded from the mode analysis. Finally, the interaction of the in-plane electric field of the guided modes with graphene induces optical absorption as a result of the generation of a surface current. The absorption coefficient in dB per unit length, $\alpha$, can be expressed as

\begin{equation}
    \alpha=10(\log_{10} \mathrm{e})\frac{\mathrm{Re}(\sigma_g)}{2 P_{in}} \int_{w_g} |E_{\parallel}|^2,
\end{equation}

where $E_{\parallel}$ corresponds to the in-plane electric field that interacts with graphene and $w_g$ is the width of the nanoribbons \cite{koester_waveguidecoupled_2014a}. The remaining materials considered in this work (Si$_3$N$_4$, Si, SiO$_2$ and Al$_2$O$_3$) are transparent at the operating wavelengths and therefore do not induce any optical absorption.

An alternative approach including the complete mode analysis of the hybrid waveguide was also explored. Graphene was modeled as a 1 nm thick layer with a permittivity of $\varepsilon_g =1+(i \sigma_g)/(\omega \varepsilon_0 d)$. However, it resulted in a computationally intensive simulation with a relative offset of less than 10\% in the propagation losses. Since the inclusion of graphene in the analysis of the mode distribution did not affect the identification of the optimal nanoribbon geometries, this second method was disregarded.

\subsection{Selective modulation structure}

In order to maintain a high mode selectivity, one requirement of the mode-selective modulators is that they can provide a high extinction and selection ratio between the target modulated mode in its attenuated state and the non-modulated mode. If the absorption coefficient in dB per unit length is designated as $\alpha_{m}$ for the modulated mode in its high-loss state and $\alpha_{p}$ for the remaining non-modulated propagating mode, the extinction and selection ratio can be defined as $ER=|\alpha_m|-|\alpha_p|$ and $SR=ER/|\alpha_p|$ \cite{xing_ultrabroadband_2018a, martin-romero_design_2026b}. These parameters are related to the propagation distance and insertion losses required to achieve a contrast of $C$ in dB units as $L=C/ER$ and $|IL|=C/SR$, respectively. Thus, the optimal configuration that minimizes the product $L\times |IL|=C^2/(ER\times SR)$ is the one that maximizes $ER\times SR$.

\begin{figure}[ht!]
\centering
\includegraphics[width=\linewidth]{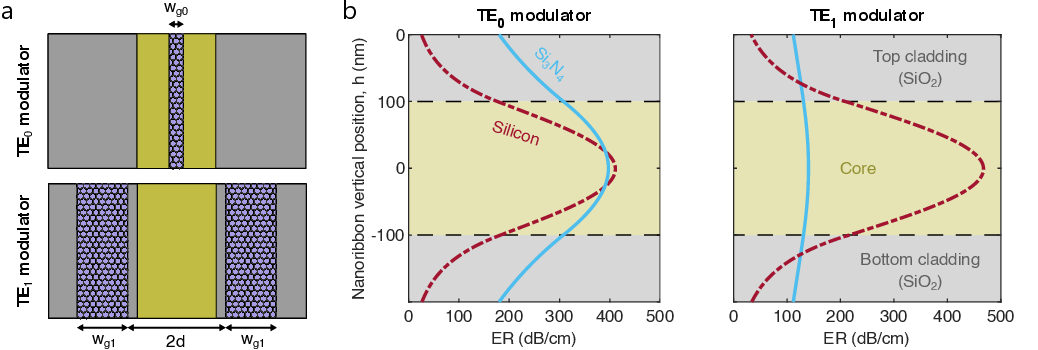}
\caption{(a) Schematic of the graphene nanoribbons in the TE$_0$ and TE$_1$ modulators, with depicted optimization parameters. (b) Extinction ratio of both graphene-silicon and graphene-Si$_3$N$_4$ modulators when the nanoribbons are integrated at varying layer depths. The waveguide cross sections are 1500 nm $\times$ 200 nm and 1850 nm $\times$ 200 nm for silicon and Si$_3$N$_4$, respectively. The corresponding nanoribbon geometrical parameters are $w_{g0}=260$ nm, $w_{g1}=1000$ nm, $d=400$ nm (silicon) and $w_{g0}=500$ nm, $w_{g1}=1400$ nm, $d=1000$ nm (Si$_3$N$_4$). Dashed lines indicate the interfaces between the top cladding and core layers, and between the core and bottom cladding layers.}
\label{fig:2}
\end{figure}

The fixed dimensions selected for the waveguide core in graphene-Si$_3$N$_4$ modulators were 1850 nm $\times$ 200 nm. The low thickness provides a stronger evanescent field at the position of the nanoribbons, while the core width ensures that the waveguide only supports the first two modes for transverse electrically polarized light. The product of $ER\times SR$ was optimized in each mode-selective modulator for a wavelength of 1550 nm by simulating the absorption coefficient of the TE$_0$ and TE$_1$ modes while varying the geometric parameters of the nanoribbons \cite{martin-romero_design_2026b}. The resulting extinction and selection ratio were compared to those reported for graphene-silicon with 1500 nm $\times$ 200 nm core dimensions \cite{xing_ultrabroadband_2018a}. These values are summarized in table \ref{tab:1}, while the explored geometric parameters of the nanoribbons are depicted schematically in figure \ref{fig:2}(a). From a physical perspective, the reported nanoribbon geometry that maximizes $ER\times SR$ allows to maintain a high overlap between the field distribution of each guided mode and its corresponding mode-selective modulator, while concurrently maintaining a low overlap with the mode-selective modulator corresponding to the other guided mode. This results in a central nanoribbon that modulates the TE$_0$ mode by overlapping with its central intensity maximum, and two side nanoribbons that modulate the TE$_1$ mode by overlapping with its pronounced evanescent side tails. It can be observed that configurations based on graphene-Si$_3$N$_4$ generally produce higher mode selectivity. Specifically, the selection ratio was found to be higher for both the TE$_0$ and the TE$_1$ modulators, at the expense of a lower extinction ratio for the latter. This is due to the lower refractive index of silicon nitride compared to silicon ($n_{Si_3N_4}=2$ and $n_{Si}=3.48$ at a wavelength of $1550$ nm), which provides a weaker confinement and allows the in-plane evanescent field to interact more strongly with the graphene nanoribbons. In addition, for the TE$_1$ modulator, the nanoribbons can be positioned farther from the waveguide core because the TE$_1$ mode exhibits a deeper evanescent field penetration. This enables efficient interaction with graphene while reducing the undesired absorption losses experienced by the TE$_0$ mode.

\begin{table}[ht!]
\centering
\caption{Optimized geometry of a TE$_0$ and a TE$_1$ modulator in the graphene-silicon and graphene-Si$_3$N$_4$ platforms, for waveguides with cross sections of 1500 nm $\times$ 200 nm (silicon) and 1850 nm $\times$ 200 nm (silicon nitride), respectively. The nanoribbons are considered to be transferred at the plane of the interface between the core and top cladding layers.}
\label{tab:1}
\vspace{6pt}
\begin{tabular}{lccp{5pt}cc}
\toprule
 & \multicolumn{2}{c}{TE$_0$ modulator} & & \multicolumn{2}{c}{TE$_1$ modulator} \\
\cmidrule{2-3} \cmidrule{5-6}
Platform & Si & Si$_3$N$_4$ & & Si & Si$_3$N$_4$ \\
\midrule
w$_{g0}$ (nm) & 260 & 500  & & -    & -    \\
w$_{g1}$ (nm) & -   & -    & & 1000 & 1400 \\
d (nm)        & -   & -    & & 400  & 1000 \\
ER (dB/cm)    & 180 & \textbf{306} & & \textbf{209} & 131  \\
SR            & 3.8 & \textbf{4.4} & & 1.6  & \textbf{4.3} \\
\bottomrule
\end{tabular}
\end{table}

The optimization shown in table \ref{tab:1} was performed for mode-selective modulators comprising graphene nanoribbons transferred on top of the waveguide core, or at the height of that same plane, prior to the deposition of a silicon dioxide top cladding layer. However, the effect of the weaker confinement in Si$_3$N$_4$ becomes more apparent when the graphene nanoribbons are embedded at the layer depth of the waveguide core. This can be achieved through additional fabrication steps, by depositing a new amorphous silicon or silicon nitride layer on top of the waveguide core, after transferring the nanoribbons and before the deposition of the top cladding. Figure \ref{fig:2}(b) shows the simulated extinction ratio of the TE$_0$ and the TE$_1$ modulators for a varying vertical position within the core thickness, defined relative to its mid-plane layer. Comparing both material platforms, the results indicate that the graphene-silicon  modulators achieve a significantly higher extinction ratio when the nanoribbons are embedded exactly at the mid-plane layer of the core, where the intensity of the electric field is stronger. In contrast, this difference is less pronounced in the case of graphene-Si$_3$N$_4$ modulators, where the lower refractive index implies a slow decay rate in the evanescent field. It can be concluded that the silicon nitride platform is more suitable for a design in which graphene interacts with the evanescent field at the core-cladding interface. Such a configuration avoids embedding the nanoribbons within the core layer, thereby simplifying the fabrication process.

Regarding the viability of the fabrication, the tolerance to potential deviations in the geometry of the nanoribbons was also analyzed. The tolerance range was defined as a confidence interval in which the extinction ratio remains above 90\% of its optimized value. The primary source of deviations that can arise during the fabrication of the TE$_0$ modulator is the misalignment of the central nanoribbon with respect to the symmetry plane of the waveguide core. The tolerance range for this misalignment was found to be 80 nm for a silicon waveguide and 150 nm for a Si$_3$N$_4$ waveguide, corresponding to a misalignment of 30.77\% and 30\% relative to the nanoribbon width, respectively. In the case of the TE$_1$ modulator, the most significant source of fabrication errors lies on a vertical offset in the graphene nanoribbons that can be totally or partially located on the side claddings. The tolerance to this offset is only 10 nm for a silicon waveguide, while it goes up to 70 nm for a Si$_3$N$_4$ waveguide. Thus, silicon nitride offers greater fabrication robustness for graphene mode-selective modulators.

\begin{figure}[ht!]
\centering
\includegraphics[width=\linewidth]{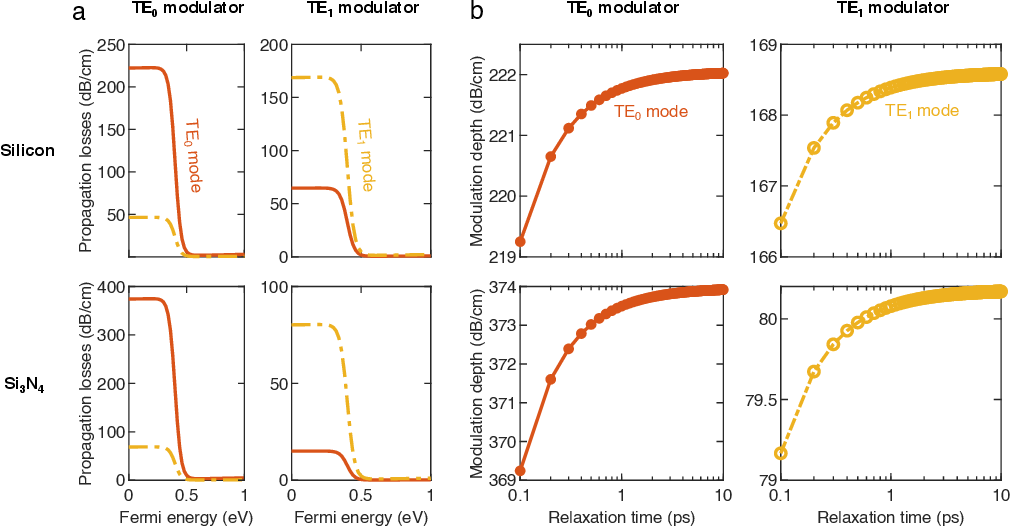}
\caption{(a) Propagation losses of the TE$_0$ (solid line) and TE$_1$ (dashed line) modes as a function of the Fermi energy of graphene, simulated for TE$_0$ or TE$_1$ modulators on top of the core of silicon or silicon nitride waveguides. (b) Modulation depth as a function of charge relaxation time in graphene, for TE$_0$ or TE$_1$ modulators on top of the core of silicon or silicon nitride waveguides. The waveguide cross sections are 1500 nm $\times$ 200 nm and 1850 nm $\times$ 200 nm for silicon and Si$_3$N$_4$, respectively. The corresponding nanoribbon geometrical parameters are $w_{g0}=260$ nm, $w_{g1}=1000$ nm, $d=400$ nm (silicon) and $w_{g0}=500$ nm, $w_{g1}=1400$ nm, $d=1000$ nm (Si$_3$N$_4$). The operating wavelength is $1550$ nm.}
\label{fig:3}
\end{figure}

The efficiency of active modulation in the mode-selective devices was evaluated by tuning the Fermi energy of graphene in the simulations. Figure \ref{fig:3}(a) shows the propagation losses of the TE$_0$ and TE$_1$ modes as a function of $E_f$ for both types of nanoribbon designs (TE$_0$ modulator or TE$_1$ modulator) and in both material platforms (silicon or Si$_3$N$_4$). As expected, the propagation losses were effectively suppressed for $E_f\gtrapprox0.4$ eV. Owing to the optimized geometry of the nanoribbons, the opaque to transparent transition is more pronounced for the targeted mode in each of the mode-selective modulator configurations.

The modulation depth was defined as the contrast in propagation losses of the targeted mode between Fermi energies below and above the transition threshold, expressed in dB per unit length. Figure \ref{fig:3}(b) shows the simulated modulation depth as a function of the charge relaxation time. Because spurious optical absorption due to intraband processes in the transparent regime decreases for longer relaxation times, the modulation depth of the nanoribbons is correspondingly enhanced. The results indicate that the Si$_3$N$_4$ platform provides greater modulation depth than silicon in the TE$_0$ modulators, while the opposite trend is observed for TE$_1$ modulators. However, the observed dependence on the charge relaxation time is weak over the considered range, indicating a low sensitivity of the modulation performance to carrier scattering dynamics.

\section{Dual-mode modulator}

\subsection{Simultaneous modulation structure}

We present the concept and design of a graphene-Si$_3$N$_4$ dual-mode modulator. It combines the TE$_0$ and TE$_1$ mode-selective modulators in one waveguide. Silicon nitride was selected as the preferred platform because it offers overall better performance, as discussed in the previous section. Although silicon provides a higher modulation depth and extinction ratio for the TE$_1$ modulator, Si$_3$N$_4$ shows superior values for the TE$_0$ modulator and achieves a higher selection ratio in both devices. It is able to maintain a high mode selectivity, preserving the distinguishability when both modes are modulated simultaneously, and it is more robust against structural deviations in the fabrication process.

The proposed dual-mode modulator allows to switch the absorption of the TE$_0$ and TE$_1$ modes simultaneously and independently, by modifying the Fermi energy of the central nanoribbon or the side nanoribbons, respectively. Four possible logical states are achieved, denoted as (0,0), (0,1), (1,0) and (1,1). They correspond to a strong absorption of both modes, the TE$_0$ mode, the TE$_1$ mode or none of them. Although we optimized the geometry of the TE$_0$ and TE$_1$ modulators individually as tunable filters, the dual-mode modulator requires a dedicated optimization process because it must integrate both functionalities simultaneously. This additional optimization ensures that all switching operations between the four joint logical states can be performed efficiently. Table \ref{tab:2} summarizes the main parameters corresponding to three of the different designs explored for the device. The (0,0)$\leftrightarrow$(1,1) and (0,1)$\leftrightarrow$(1,0) state transitions produce the maximum and minimum modulation depth, respectively. Among the three cases, the selected design corresponds to a  width of $w_{g0}=200$ nm for the central nanoribbon, as well as a width of $w_{g1}=1600$ nm and a horizontal offset of $d=1100$ nm for the side nanoribbons. It was chosen for two main reasons. First, the limiting factor in the device performance, which ultimately determines its footprint, is the modulation depth of the TE$_1$ mode between the (1,0) and (0,1) states. This design provides the highest value. The second reason is that it also maintains the highest selection ratio for both modes, ensuring a high mode selectivity.

\begin{table}[ht!]
\centering
\caption{Modulation depth, extinction and selection ratio for three different monolayer graphene-Si$_3$N$_4$ combined TE$_0$ and TE$_1$ modulator geometries in a waveguide with a 1850 nm × 200 nm cross section for a wavelength of $1550$ nm.}
\label{tab:2}
\vspace{6pt}
\resizebox{\textwidth}{!}{%
\begin{tabular}{cccp{5pt}ccccp{5pt}cccc}
\toprule
 & & & & \multicolumn{4}{c}{TE$_0$ mode} & & \multicolumn{4}{c}{TE$_1$ mode} \\
\cmidrule{5-8} \cmidrule{10-13}
\makecell{w$_{g0}$\\(nm)} & \makecell{w$_{g1}$\\(nm)} & \makecell{d\\(nm)} & &
\makecell{MD\\(dB/cm)\\($\underline{\textbf{0}}$,0)$\leftrightarrow$($\underline{\textbf{1}}$,1)} &
\makecell{MD\\(dB/cm)\\($\underline{\textbf{0}}$,1)$\leftrightarrow$($\underline{\textbf{1}}$,0)} &
\makecell{ER\\(dB/cm)\\($\underline{\textbf{0}}$,1)} &
\makecell{SR\\($\underline{\textbf{0}}$,1)} & &
\makecell{MD\\(dB/cm)\\(0,$\underline{\textbf{0}}$)$\leftrightarrow$(1,$\underline{\textbf{1}}$)} &
\makecell{MD\\(dB/cm)\\(0,$\underline{\textbf{1}}$)$\leftrightarrow$(1,$\underline{\textbf{0}}$)} &
\makecell{ER\\(dB/cm)\\(1,$\underline{\textbf{0}}$)} &
\makecell{SR\\(1,$\underline{\textbf{0}}$)} \\
\midrule
700 & 1600 & 1000 & & 524 & 462 & 379 & 3.1 & & 281 & 44 & 128 & 3.1 \\
500 & 1400 & 1000 & & 393 & 332 & 302 & 4.3 & & 227 & 95 & 125 & 3.2 \\
\textbf{200} & \textbf{1600} & \textbf{1100} & & 167 & 135 & 133 & \textbf{6.4} & & 132 & \textbf{95} & 97 & \textbf{5.2} \\
\bottomrule
\end{tabular}%
}
\end{table}

\begin{figure}[ht!]
\centering
\includegraphics[width=\linewidth]{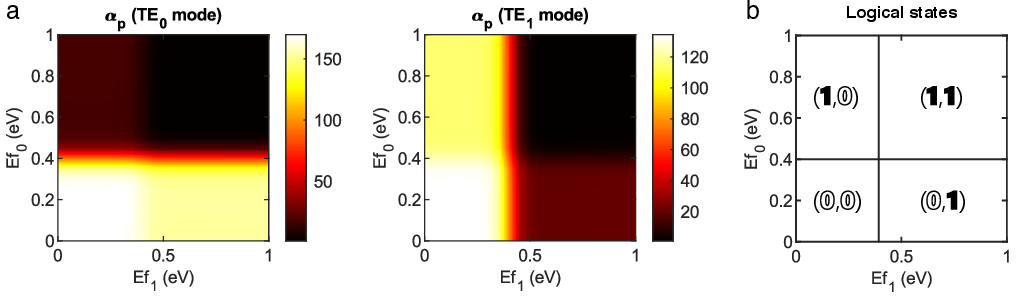}
\caption{(a) Absorption coefficient of the TE$_0$ and TE$_1$ modes as a function of the Fermi energy of the TE$_0$ and TE$_1$ modulators in an optimized single layer graphene-Si$_3$N$_4$ nanoribbon geometry ($w_{g0}=200$ nm, $w_{g1}=1600$ nm, $d=1100$ nm). (b) This geometry leads to switching between four possible logical states: (0,0), (0,1), (1,0) and (1,1). The waveguide core dimensions are 1850 nm $\times$ 200 nm, and the wavelength is $1550$ nm.}
\label{fig:4}
\end{figure}

Figure \ref{fig:4}(a) presents the propagation losses of the TE$_0$ and TE$_1$ modes in the configuration corresponding to the selected design, as a function of the Fermi energy of graphene in the TE$_0$ modulator ($E_{f0}$) and the TE$_1$ modulator ($E_{f1}$). It is verified that the switching of the optical absorption in each of these two modes depends mainly on the opaque to transparent transition of their corresponding mode-selective modulator. Four regions can be identified as shown in figure \ref{fig:4}(b). When $E_{f0}$ and $E_{f1}$ exceed the threshold of 0.4 eV, both modulators enter the transparent regime, and the (1,1) state is reached. When they are below the threshold, both modes are absorbed and the (0,0) state is reached. If only one of the mode-selective modulators meets the condition $E_f>0.4$ eV, then the targeted mode is transmitted while the other is absorbed, generating the (1,0) or (0,1) states. Mode selectivity becomes a critical parameter when the device operates in these last two states.

\subsection{Dual-layer graphene switching}

Tuning of the Fermi energy (or electrochemical potential) of graphene can be achieved through electrical gating. To apply a gate voltage to graphene, a nanoribbon capacitor structure is proposed. Instead of using a single graphene monolayer, the final structure consists of two monolayers separated by an Al$_2$O$_3$ layer that serves as a dielectric spacer. Each monolayer is connected to transparent conductive oxide (TCO) electrodes, through which the voltage is applied (figure \ref{fig:1}(c)). The charge density, $\rho$, in the top and bottom monolayer is, respectively \cite{koester_highspeed_2012},

\begin{gather}
    \rho_{top}=-e n_b+\frac{1}{A}\int\displaylimits_{\Delta V_g} C \mathrm{d}V, \\
    \rho_{bot}=-e n_b-\frac{1}{A}\int\displaylimits_{\Delta V_g} C \mathrm{d}V,
\end{gather}

where $e$ is the electron charge, $n_b$ is the natural carrier density of graphene, $A$ is the capacitor area, $\Delta V_g$ is the applied gate voltage, and $C$ is the total capacitance. If natural doping is negligible, the top and bottom charge densities satisfy the relation $\rho_{top}=-\rho_{bot}$, and therefore the Fermi energy of both layers is $E_{f;top}=-E_{f;bot}$. In that case, both layers serve as gate contacts and modulated absorbers \cite{liu_doublelayer_2012a}. However, natural doping may also be increased to achieve switching from a high charge density to $\rho=0$ in either the top or bottom layer. The Fermi energy of that layer would undergo a transparent to opaque transition, while the other would stay in a transparent state. Therefore, one layer would serve only as a gate contact, while the other would induce modulated absorption \cite{koester_highspeed_2012}. In the present work, we have focused on the naturally undoped case.

From a fabrication perspective, the proposed device could be implemented following a process analogous to that experimentally demonstrated for graphene–Si$_3$N$_4$ mode filters \cite{martin-romero_design_2026b}. The fabrication would first involve SiO$_2$ deposition and planarization over the Si$_3$N$_4$ waveguide, followed by the transfer of the graphene monolayers and atomic layer deposition of Al$_2$O$_3$. The resulting graphene–Al$_2$O$_3$–graphene stack would then be patterned to define the nanoribbon capacitors. Finally, the electrical contacts would be defined and the upper SiO$_2$ layer would be deposited.

\begin{figure}[ht!]
\centering
\includegraphics[width=\linewidth]{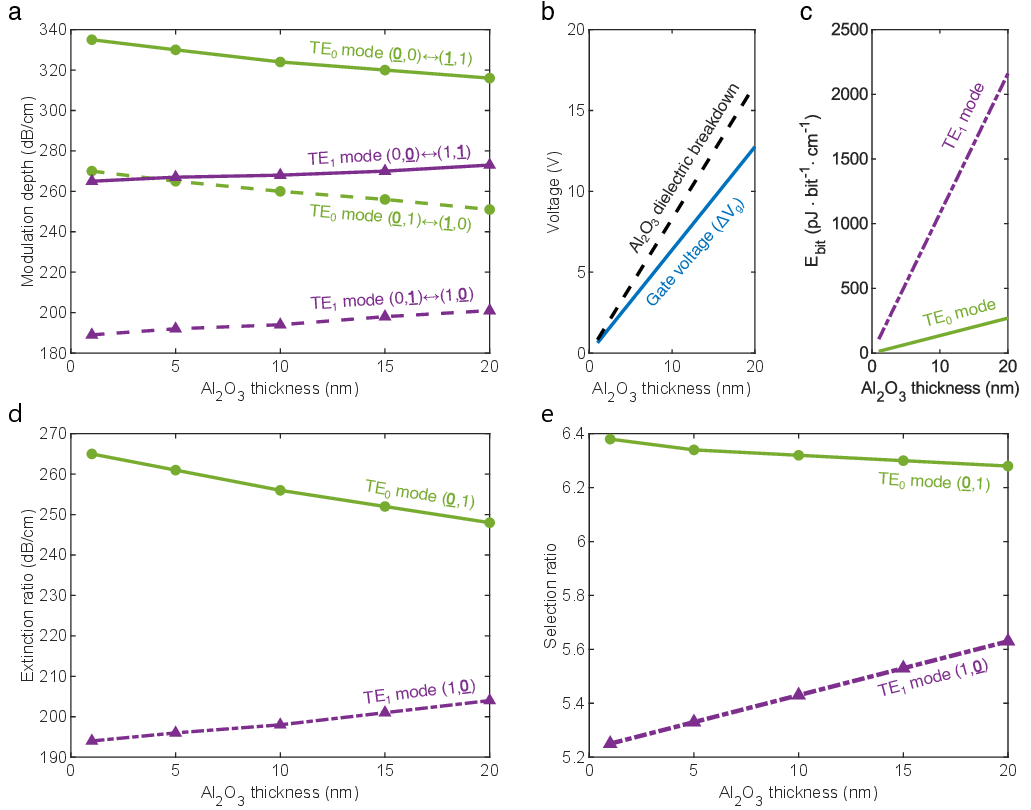}
\caption{(a) Maximum and minimum modulation depth for the TE$_0$ and TE$_1$ modes in dual-layer graphene-silicon nitride dual-mode modulators with different Al$_2$O$_3$ thicknesses (1--20 nm). (b) Gate voltage required for switching the TE$_0$ and TE$_1$ modes. (c) Consumed energy per bit and unit length for switching the TE$_0$ and TE$_1$ modes. (d) Extinction ratio (ER) for the filtering of the TE$_0$ and TE$_1$ modes in the (0,1) and (1,0) logical states, respectively. (e) Selection ratio (SR) for the filtering of both modes in the same states. The geometrical parameters for the 1850 nm $\times$ 200 nm waveguide core are $w_{g0}=200$ nm, $w_{g1}=1600$ nm, $d=1100$ nm, and the operating wavelength is $1550$ nm.}
\label{fig:5}
\end{figure}

Figure \ref{fig:5}(a) shows the simulated maximum and minimum modulation depth of a naturally undoped graphene dual-mode modulator for both modes, as a function of the spacer thickness in the nanoribbon capacitors. The optical absorption of the upper and lower graphene nanoribbons was calculated independently by considering their respective positions within the modulator structure, and the effect of the dielectric spacer on the mode profiles. The total optical absorption in dB per unit length was obtained for the TE$_0$ and TE$_1$ modes by summing the contributions of both graphene layers. The addition of the second layer of graphene was observed to enhance optical absorption. Since the conductivity of graphene does not depend on the sign of $E_f$, the resulting modulation depth was found to be approximately duplicated, owing to the condition $E_{f;top}=-E_{f;bot}$. Nevertheless, a dependence on the Al$_2$O$_3$ spacer thickness can still be observed due to the modification of the modal field distributions.

Figure \ref{fig:5}(b) displays the minimum gate voltage required to achieve active switching, which was computed as

\begin{equation}
  \Delta V_g=\frac{E_f^2}{\hbar^2 v_f^2 \pi \eta},  
\end{equation}

where $v_F$=1$\times$10$^6$ m/s is the Fermi velocity in graphene \cite{bonaccorso_graphene_2010,koester_highspeed_2012} and $\eta=\varepsilon_0 \varepsilon_r/(t e)$ is given by a plane-capacitor model of the nanoribbons \cite{shu_significantly_2018}. A representative static relative permittivity of $\varepsilon_r=7.5$ was adopted for the Al$_2$O$_3$ spacer, consistent with the experimentally reported values for amorphous Al$_2$O$_3$ grown through atomic layer deposition (ALD) \cite{kim_influence_2022}. The required voltage is below the dielectric breakdown voltage of Al$_2$O$_3$ \cite{kim_influence_2022}, also depicted in the figure. Taking into account that the capacitance per unit length of the nanoribbons is expressed as $C=\varepsilon_r \varepsilon_0 w_g/t$, where $t$ is the spacer thickness, then the energy consumed per bit and unit length was calculated as \cite{watts_vertical_2011}

\begin{equation}
    E_{bit}=C (\Delta V_g^2/4),
\end{equation}

and is presented in figure \ref{fig:5}(c).

When the dual-layer modulator operates in the logical states (1,0) or (0,1), a high contrast between the absorbed and transmitted mode is reached, ensuring high mode selectivity. The extinction and selection ratio for both states is shown in figure \ref{fig:5}(d-e). Taking into account that the optical performance of the device is limited by the operation on the TE$_1$ mode, the optimal thickness for the dielectric spacer can be established as $t=20$ nm. The corresponding simulated device performance consists of a modulation depth of 251--316 dB/cm for the TE$_0$ mode and 201--273 dB/cm for the TE$_1$ mode, with an extinction ratio of 248 dB/cm for the TE$_0$ mode and 204 dB/cm for the TE$_1$ mode, and a selection ratio of 6.28 for the TE$_0$ mode and 5.63 for the TE$_1$ mode. However, the optical performance exhibits a considerably weaker dependence on the dielectric spacer thickness than the energy consumption. Reducing $t$ from $20$ to $5$ nm decreases the energy consumption from 270 and 2160 pJ bit$^{-1}$ cm$^{-1}$ to 67.5 and 540 pJ bit$^{-1}$ cm$^{-1}$ for the TE$_0$ and TE$_1$ modes, respectively, while only marginally reducing the optical performance of the limiting TE$_1$ modulator and slightly improving that of the TE$_0$ modulator. Thus, smaller dielectric thicknesses can provide a favorable trade-off between optical performance and energy consumption.

The performance parameters reported per unit length can be directly translated into the corresponding values for a finite-length device. Specifically, for a target modal contrast in dB, $C$, in the (0,1) or (1,0) states, the total active length can be obtained from the extinction ratio as $L=C/ER$. The corresponding unwanted insertion losses of the transmitted mode in these states will be higher than in the (1,1) state, and can be derived from the selection ratio as $|IL|=C/SR$. The modulation depth and energy consumption can be readily multiplied by the active length of the dual-mode modulator. Representative finite-length waveguides targeting extinction ratios of 3, 6, and 10 dB for the limiting TE$_1$ mode are summarized in table \ref{tab:3}.

\begin{table}[ht!]
\centering
\caption{Total optical performance and energy consumption for three different waveguide and nanoribbon active lengths in the dual-layer dual-mode modulator, at a wavelength of $1550$ nm. The displayed configurations correspond to target modal contrasts of 3, 6, and 10 dB for the limiting TE$_1$ mode. The geometrical parameters for the 1850 nm $\times$ 200 nm waveguide core are $w_{g0}=200$ nm, $w_{g1}=1600$ nm, $d=1100$ nm. Two Al$_2$O$_3$ spacer thicknesses are inspected: $20$ nm and $5$ nm.}
\label{tab:3}
\vspace{6pt}
\resizebox{\columnwidth}{!}{%
\begin{tabular}{llp{5pt}cccccp{5pt}ccccc}
\toprule
& & & \multicolumn{5}{c}{TE$_0$ modulation} & & \multicolumn{5}{c}{TE$_1$ modulation} \\
\cmidrule{4-8} \cmidrule{10-14}
\makecell{Al$_2$O$_3$\\thickness\\(nm)} & \makecell{Length\\($\mathrm{\upmu}$m)} & &
\makecell{MD\\(dB)\\($\underline{\textbf{0}}$,0)$\leftrightarrow$($\underline{\textbf{1}}$,1)} &
\makecell{MD\\(dB)\\($\underline{\textbf{0}}$,1)$\leftrightarrow$($\underline{\textbf{1}}$,0)} &
\makecell{C\\(dB)\\($\underline{\textbf{0}}$,1)} &
\makecell{$|$IL$|$\\(dB)\\($\underline{\textbf{0}}$,1)} &
\makecell{E$_{bit}$\\(pJ/bit)} & &
\makecell{MD\\(dB)\\(0,$\underline{\textbf{0}}$)$\leftrightarrow$(1,$\underline{\textbf{1}}$)} &
\makecell{MD\\(dB)\\(0,$\underline{\textbf{1}}$)$\leftrightarrow$(1,$\underline{\textbf{0}}$)} &
\makecell{C\\(dB)\\(1,$\underline{\textbf{0}}$)} &
\makecell{$|$IL$|$\\(dB)\\(1,$\underline{\textbf{0}}$)} &
\makecell{E$_{bit}$\\(pJ/bit)} \\
\midrule
\multirow{3}{*}{20} & 147 & & 4.65  & 3.69  & 3.65  & 0.58 & 3.97  & & 4.01  & 2.95 & \textbf{3}  & 0.53 & 31.75 \\
                     & 294 & & 9.29  & 7.38  & 7.29  & 1.16 & 7.94  & & 8.03  & 5.91 & \textbf{6}  & 1.07 & 63.50 \\
                     & 490 & & 15.48 & 12.30 & 12.15 & 1.93 & 13.23 & & 13.38 & 9.85 & \textbf{10} & 1.78 & 105.84 \\
\midrule
\multirow{3}{*}{5}  & 153 & & 5.05  & 4.05  & 3.99  & 0.63 & 1.03 & & 4.09  & 2.94 & \textbf{3}  & 0.57 & 8.26 \\
                     & 306 & & 10.10 & 8.11  & 7.99  & 1.25 & 2.07 & & 8.17  & 5.88 & \textbf{6}  & 1.14 & 16.52 \\
                     & 510 & & 16.83 & 13.52 & 13.31 & 2.09 & 3.44 & & 13.62 & 9.79 & \textbf{10} & 1.90 & 27.54 \\
\bottomrule
\end{tabular}%
}
\end{table}

\begin{figure}[ht!]
\centering
\includegraphics[width=\linewidth]{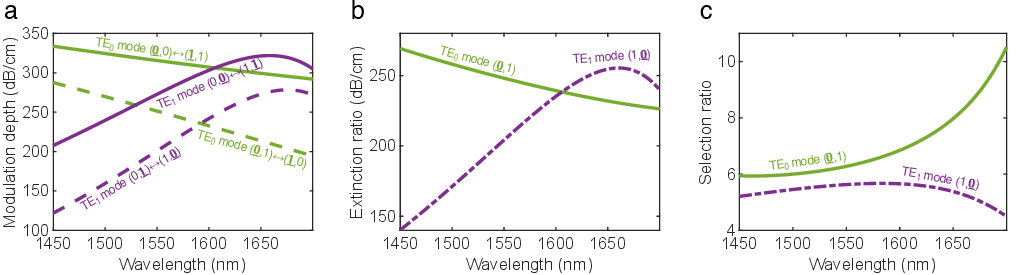}
\caption{(a) Maximum and minimum modulation depth for the TE$_0$ and TE$_1$ modes in dual-layer graphene-silicon nitride dual-mode modulators as a function of wavelength. (b) Spectral dependence of the extinction ratio (ER) for the filtering of the TE$_0$ and TE$_1$ modes in the (0,1) and (1,0) logical states, respectively. (c) Spectral dependence of the selection ratio (SR) for the filtering of both modes in the same states. The geometrical parameters for the 1850 nm $\times$ 200 nm waveguide core are $w_{g0}=200$ nm, $w_{g1}=1600$ nm, $d=1100$ nm. The Al$_2$O$_3$ spacer thickness is $20$ nm.}
\label{fig:6}
\end{figure}

The wavelength-dependent performance of the proposed dual-mode modulator for an Al$_2$O$_3$ thickness of $t=20$ nm is presented in figure \ref{fig:6}. The maximum and minimum modulation depth achieved for each mode in the (0,0)$\leftrightarrow$(1,1) and (0,1)$\leftrightarrow$(1,0) transitions as a function of wavelength are shown in figure \ref{fig:6}(a). The spectrum of the extinction and selection ratio for the TE$_0$ and TE$_1$ modes in the (0,1) and (1,0) states are displayed in figure \ref{fig:6}(b-c). Although graphene exhibits intrinsically broadband and homogeneous optical properties in the considered spectral range, the spatial distribution of the electric field in the TE$_0$ and TE$_1$ modes varies with the operating wavelength. This change in the modal profiles modifies their overlap with the graphene nanoribbons and, consequently, the absorption experienced by each mode. Therefore, the inherently non-resonant behavior resulting from the waveguide architecture of the modulator and the homogeneous spectral properties of graphene exhibits smooth spectral features arising from the variation of the overlap between the guided modes and the nanoribbons as a function of wavelength. Within the investigated wavelength range, the waveguide maintains the two TE modes considered throughout this work, with no additional TE$_2$ mode or cutoff of the TE$_1$ mode.

Finally, the modulation speed of a practical implementation is expected to be defined by the RC response of the electrical driving architecture. The theoretical operating speed limits of graphene electro-absorption modulators are in the hundreds of GHz, and they are determined by electrical limitations such as the graphene sheet resistance rather than by its ultrafast intrinsic carrier dynamics \cite{gosciniak_theoretical_2013}. Therefore, the corresponding electrical bandwidth depends on the TCO electrode resistance, graphene-TCO contact resistance, graphene sheet resistance, electrode geometry and configuration, interconnect parasitic resistances, and driver impedance, in addition to the capacitance of the graphene-Al$_2$O$_3$-graphene nanoribbons. Reducing the mentioned electrical resistances is desirable to achieve a higher electrical bandwidth. The use of a TCO material is particularly advantageous in the proposed configuration in order to avoid an additional optical absorption introduced by the electrodes. Their geometry should provide a trade-off, since increasing the graphene-TCO interface area can reduce their total contact resistance, but at the expense of introducing electrode-induced absorption losses. Moreover, the different widths of the central and side nanoribbons result in different capacitances and may also lead to different electrical resistances, potentially yielding different RC bandwidth limitations for the TE$_0$ and TE$_1$ modulators. Depending on the contact configuration, distributed RC effects may also become relevant. Overall, these considerations indicate that the quantitative assessment of the modulation speed would require the definition of a specific electrical contact and driving architecture, which goes beyond the scope of this work.

\section{Conclusion}

We have reported the design and simulation of an integrated graphene-silicon nitride dual-mode modulator presenting high mode selectivity. The device allows to achieve simultaneous selective electro-absorption modulation for the TE$_0$ and TE$_1$ modes in a buried dual-mode waveguide, switching between four logical states: (0,0), (0,1), (1,0) and (1,1). These states are defined as the strong absorption of both modes, the TE$_0$ mode, the TE$_1$ mode or none of them. The working principle of the active modulation relies on the electrical tuning of the Fermi energy levels in three graphene-Al$_2$O$_3$-graphene capacitor nanoribbons, and their evanescent interaction with the guided modes.

In order to reach an optimal configuration for the dual-mode modulator, we first studied the mode-selective modulation in single-layer graphene nanoribbons. As a result, we established Si$_3$N$_4$-on-insulator as the preferred platform for highly mode selective evanescent interaction, owing to a higher fabrication robustness and a better overall performance due to the lower refractive index contrast in the core-cladding interface. Then, we combined both mode-selective modulators and optimized the nanoribbon geometrical parameters to reach efficient simultaneous modulation of the TE$_0$ and TE$_1$ modes. Finally, we studied the performance of the dual-mode modulator for graphene-Al$_2$O$_3$-graphene capacitor nanoribbons with the same geometry.

The optically optimized design consists of a 1850 nm $\times$ 200 nm Si$_3$N$_4$-SiO$_2$ embedded dual-mode waveguide with a 200 nm wide central capacitor and two 1600 nm wide side capacitors, separated 1100 nm from the waveguide symmetry plane. If the thickness of the dielectric spacer layer located between both graphene monolayers in each nanoribbon is 20 nm, the simulated modulation depth is 251--316 dB/cm for the TE$_0$ mode and 201--273 dB/cm for the TE$_1$ mode, respectively. The simulated extinction ratio and selection ratio of the TE$_0$ mode corresponding to the (0,1) state are 248 dB/cm and 6.28. The simulated extinction and selection ratio of the TE$_1$ mode in the (1,0) state are 204 dB/cm and 5.63. This configuration would yield a modulation voltage of 12.8 V and an energy consumption per unit length of 270 and 2160 pJ bit$^{-1}$ cm$^{-1}$ for the TE$_0$ and TE$_1$ modes. However, flexibility in the design is a strong advantage of this device, since the energy consumption can be significantly reduced at the expense of a marginally lower optical performance of the TE$_1$ modulator by selecting a thinner dielectric layer for the nanoribbons, as can be observed in figure \ref{fig:5} and table \ref{tab:3}. For instance, an Al$_2$O$_3$ thickness of 5 nm would yield a gate voltage of 3.2 V and a consumption of 67.5 and 540 pJ bit$^{-1}$ cm$^{-1}$ for the TE$_0$ and TE$_1$ modes. Under this configuration, the modulation depth of each mode would be 265--330 dB/cm and 192--267 dB/cm, respectively. The extinction and selection ratio for the TE$_0$ mode in the (0,1) state would be 261 dB/cm and 6.38, while for the TE$_1$ mode in the (1,0) state they would be 196 dB/cm and 5.25, respectively. Therefore, a reduced capacitor thickness of 5 nm can be readily selected for applications that require a lower voltage and energy consumption.

In conclusion, the proposed graphene-Si$_3$N$_4$ dual-mode waveguide allows simultaneous modulation of two different guided modes with high modulation depth and mode selectivity. Table \ref{tab:4} compares the simulated performance of our design with representative state-of-the-art multimode modulators reported in the literature.

\begin{table}[ht!]
\centering
\caption{Comparison of the performance of our simulated dual-layer graphene-Si$_3$N$_4$ dual-mode modulator at a wavelength of $1550$ nm with respect to the state of the art. For our proposed device, Al$_2$O$_3$ thicknesses of $20$ nm or $5$ nm are considered, and the performance parameters are reported per unit length because the active length is a free design parameter.\\
*Estimate from provided results.\\
**Data not provided.}
\label{tab:4}
\vspace{6pt}
\resizebox{\columnwidth}{!}{%
\begin{tabular}{llllllllll}
\toprule
 & & Modes & Geometry & Footprint & \makecell[l]{Modulation\\depth} & \makecell[l]{Extinction ratio\\or modal contrast} & Selection ratio & Insertion losses & Energy consumption \\
\midrule
\multirow{2}{*}{\makecell[l]{Our design\\(Graphene-\\Si$_3$N$_4$\\waveguide\\modulator\\simulation)}} &
  \makecell[l]{$t=20$ nm} &
  \multirow{2}{*}[-2em]{2} &
  \multirow{2}{*}[-2em]{\makecell[l]{Waveguide\\(one arm)}} &
  \multirow{2}{*}[-2em]{\makecell[l]{Width: $\sim$5.4 $\mathrm{\upmu m}$\\Length: free parameter}} &
  \makecell[l]{251-316 dB/cm\\(TE$_0$)\\201-273 dB/cm\\(TE$_1$)} &
  \makecell[l]{248 dB/cm\\(TE$_0$), (0,1)\\204 dB/cm\\(TE$_1$), (1,0)} &
  \makecell[l]{6.28\\(TE$_0$), (0,1)\\5.63\\(TE$_1$), (1,0)} &
  $-$ &
  \makecell[l]{270 pJ bit$^{-1}$ cm$^{-1}$\\(TE$_0$)\\2160 pJ bit$^{-1}$ cm$^{-1}$\\(TE$_1$)} \\
\cmidrule(lr){2-2}\cmidrule(lr){6-10}
 &
  \makecell[l]{$t=5$ nm} &
   &
   &
   &
  \makecell[l]{265-330 dB/cm\\(TE$_0$)\\192-267 dB/cm\\(TE$_1$)} &
  \makecell[l]{261 dB/cm\\(TE$_0$), (0,1)\\196 dB/cm\\(TE$_1$), (1,0)} &
  \makecell[l]{6.38\\(TE$_0$), (0,1)\\5.25\\(TE$_1$), (1,0)} &
  $-$ &
  \makecell[l]{67.5 pJ bit$^{-1}$ cm$^{-1}$\\(TE$_0$)\\540 pJ bit$^{-1}$ cm$^{-1}$\\(TE$_1$)} \\
\midrule
\multicolumn{2}{l}{\makecell[l]{Graphene-polymer\\waveguide modulator\\(simulation)\\\cite{lian_modeselective_2022a}}} &
  3 &
  \makecell[l]{Waveguide\\(one arm)} &
  \makecell[l]{Width: $\sim$16 $\mathrm{\upmu m}$*\\Length: free parameter,\\selected 1 cm} &
  23 dB &
  \makecell[l]{$\sim$23 dB *\\(TE$_0$), (0,1,1)\\$\sim$19 dB *\\(TE$_{01}$, TE$_{10}$),\\(1,0,1) or (1,1,0)} &
  $-$ &
  $\lesssim$5 dB * &
  \makecell[l]{12.6 pJ bit$^{-1}$\\(TE$_0$)\\94.5 pJ bit$^{-1}$\\(TE$_{01}$)\\126 pJ bit$^{-1}$\\(TE$_{10}$)} \\
\midrule
\multicolumn{2}{l}{\makecell[l]{Graphene-Si$_3$N$_4$\\phase modulator\\(simulation)\\\cite{wang_design_2021}}} &
  2 &
  \makecell[l]{MZI\\(two arms)} &
  $\sim$0.5 mm$^2$* &
  \makecell[l]{19.3 dB\\(TE$_0$)\\21.7 dB\\(TE$_1$)} &
  \makecell[l]{20.7 dB\\(TE$_0$), (0,1)\\20.3 dB\\(TE$_1$), (1,0)} &
  $-$ &
  \makecell[l]{1.7-2.5 dB\\(TE$_0$)\\0.6-1.1 dB\\(TE$_1$)} &
  16 pJ bit$^{-1}$ \\
\midrule
\multicolumn{2}{l}{\makecell[l]{Graphene-silicon\\phase modulator\\(simulation)\\\cite{wang_design_2021a}}} &
  2 &
  \makecell[l]{MZI\\(two arms)} &
  $\sim$0.5 mm$^2$ &
  \makecell[l]{27.3 dB\\(TE$_0$)\\31.4 dB\\(TE$_1$)} &
  \makecell[l]{31.4 dB\\(TE$_0$), (0,1)\\32.3 dB\\(TE$_1$), (1,0)} &
  $-$ &
  \makecell[l]{1.6-6.5 dB\\(TE$_0$)\\2.3-7.2 dB\\(TE$_1$)} &
  1.365 pJ bit$^{-1}$ \\
\midrule
\multicolumn{2}{l}{\makecell[l]{Silicon ring\\modulator\\(experiment)\\\cite{liu_ultrahigh_2022,liu_silicon_2023}}} &
  2 &
  \makecell[l]{Racetrack\\resonator} &
  \makecell[l]{$\sim$24 $\mathrm{\upmu m}$*\\$\times$\\$\sim$42 $\mathrm{\upmu m}$*} &
  \makecell[l]{4.17 dB\\(TE$_0$)\\2.7 dB\\(TE$_1$)} &
  \makecell[l]{2.17 dB *\\(TE$_0$), (0,1)\\4.7 dB *\\(TE$_1$), (1,0)} &
  $-$ &
  \makecell[l]{$\sim$1.5 dB*\\(TE$_0$)\\$\sim$3.5 dB*\\(TE$_1$)} &
  2 pJ bit$^{-1}$ * \\
\midrule
\multicolumn{2}{l}{\makecell[l]{Graphene-silicon\\ring modulator\\(simulation)\\\cite{yue_graphenebased_2020a}}} &
  2 &
  \makecell[l]{Racetrack\\resonator} &
  \makecell[l]{$\sim$60 $\mathrm{\upmu m}$*\\$\times$\\$\sim$288.5 $\mathrm{\upmu m}$*} &
  \makecell[l]{14.6-14.7 dB\\(TE$_0$)\\9.2-15.3 dB\\(TE$_1$)} &
  \makecell[l]{14.7 dB\\(TE$_0$), (0,1)\\9.2 dB\\(TE$_1$), (1,0)} &
  $-$ &
  0.03-0.06 dB &
  ** \\
\bottomrule
\end{tabular}%
}
\end{table}

Among other reported hybrid integrated solutions based on graphene, the silicon nitride platform presents higher CMOS compatibility than polymer waveguides \cite{lian_modeselective_2022a}. In addition, the proposed device relies on electro-absorption rather than phase modulation, enabling operation with a single straight waveguide and avoiding Mach–Zehnder interferometer configurations with two arms, which simplifies the overall architecture and can significantly reduce the footprint of the functional optical section in the modulator \cite{wang_design_2021, wang_design_2021a}. Consequently, the active length becomes a free design parameter, and the device performance can be conveniently described through length-normalized figures of merit. The corresponding total performance can be directly obtained for any selected length as already discussed, with representative examples provided in table \ref{tab:3}. This adds flexibility to the device design, since the length of the modulator can be  adjusted to trade off modulation depth, extinction ratio, selection ratio, footprint and energy consumption depending on the target application. Previous studies have explored silicon or graphene-silicon dual-mode modulators based on ring resonators \cite{yue_graphenebased_2020a, liu_ultrahigh_2022, liu_silicon_2023}, and similarly, graphene-Si$_3$N$_4$ resonator implementations could be considered as a possible extension of this work. However, while resonant structures are more compact and can enhance modulation depth, they inherently impose a narrow spectral response. In contrast, the intrinsically broadband optical response of graphene \cite{nair_fine_2008} makes the proposed non-resonant configuration suitable for operation over a broader spectral range, ultimately determined by the wavelength-dependent modal characteristics of the complete device. Considering the above, the results presented in this work establish a promising pathway toward highly efficient, compact, and broadband integrated multimode optical communication systems.

\section*{Acknowledgements}

F.M.-R. acknowledges financial support from the Generalitat Valenciana (CIACIF/2022/188). V.J.G. acknowledges financial support from the Generalitat Valenciana (ESGENT/2024/17), and the AGENCIA ESTATAL DE INVESTIGACIÓN of Ministerio de Ciencia, Innovacion y Universidades Grant CNS2023-145093 funded by MICIU/AEI/10.13039/501100011033 and by “European Union NextGenerationEU/PRTR”. All authors acknowledge financial support from the AGENCIA ESTATAL DE INVESTIGACIÓN of Ministerio de Ciencia e Innovacion (PID2020-118855RB-I00 and PID2024-162261NB-I00). This study formed part of the Advanced Materials program and was supported by Ministerio de Ciencia e Innovacion (MCIN) with funding from European Union NextGenerationEU (PRTR-C17.I1) and by Generalitat Valenciana (MFA/2022/025).

\printbibliography

\end{document}